# Identifying DNA motifs based on match and mismatch alignment information

Jian-Jun SHU[*] and Kian Yan YONG

*School of Mechanical & Aerospace Engineering, Nanyang Technological University, 50 Nanyang Avenue, Singapore 639798*

**Abstract –** The conventional way of identifying DNA motifs, solely based on match alignment information, is susceptible to a high number of spurious sites. A novel scoring system has been introduced by taking both match and mismatch alignment information into account. The mismatch alignment information is useful to remove spurious sites encountered in DNA motif searching. As an example, a correct TATA box site in *Homo sapiens H4/g* gene has successfully been identified based on match and mismatch alignment information.

---

[*] Correspondence should be addressed to Jian-Jun SHU.
 E-mail address: mjjshu@ntu.edu.sg.

# I.    INTRODUCTION

The identification of DNA binding sites for transcription factors (motifs) is important for a complete understanding of co-regulation of gene expression, but still remains to be quite challenging to achieve.  Two approaches dominate motif-finding algorithms [1]: (1) the word-based way [2-4] that relies on exhaustive enumeration or counting frequencies and (2) the probabilistic way [5-7] that relies on optimizing a scalar-based scoring matrix [8, 9], which is visualized conveniently by a sequence logo [10].  However, both ways suffer from the problem of producing a high number of spurious sites, due to the sole consideration of match alignment information.

In this paper, a novel scoring system is introduced by incorporating the vector-based representation of DNA sequences into the scoring matrix.  In the vector-based representation of DNA sequences, the four nucleotide bases are placed at an equal distance from each other in a three-dimensional space.  This is possible by placing each point on the vertex of a tetrahedron.  Any point within the tetrahedron represents the different combinations of each nucleotide base type.  Therefore the weight distribution for each position can be replaced as a point in the tetrahedron in space using the three-dimensional coordinates.  This point is unique for the different weight distributions of DNA four nucleotide bases.  The advantage of this approach is twofold:  First, the number of indices used for representing the weight distribution at each position can be reduced from four to three in terms of the three-dimensional coordinates;  Second, the use of mathematical operators, namely the dot and cross products, can be adopted to describe the weight distribution in the three-dimensional coordinates and served as the basis of measuring the quantity of match and mismatch alignment information for each position.  A case study shows that, by using mismatch alignment information, a correct TATA box site in *Homo sapiens H4/g* gene has successfully been identified.

## II.    METHODS

### A.    Scalar-based representation of DNA nucleotide base code

In scalar-based scoring scheme, each DNA four nucleotide base (A, T, G and C) was represented by a set of numbers, such as *1*, *2*, *3* and *4*. The problem with this representation is that of unequal weightage assigned to each nucleotide base [11]. In order to assign the same weightage for each nucleotide base, a binary encoding scheme was proposed [12]. However, the binary encoding scheme is unable to accommodate the DNA four nucleotides fully. The complex, quaternion and hypercomplex numeral systems have been introduced into the encoding schemes by taking advantage of utilizing imaginary domain [13, 14].

### B.    Vector-based representation of DNA nucleotide base code

In vector-based scoring scheme, the coordinate system that is selected for vector-based representation is shown in Figure 1:

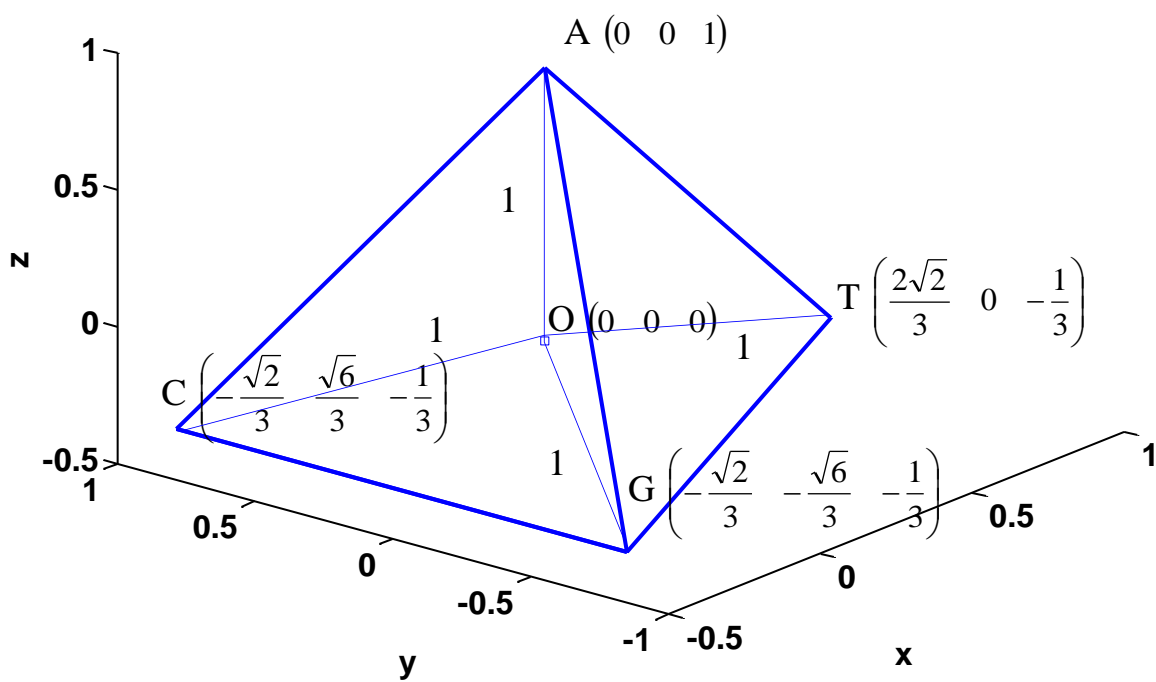

Figure 1: A tetrahedron with assigned nucleotide bases in $x$, $y$ and $z$-axes

The DNA four nucleotide bases are represented by the vertices A, T, G and C with O being the origin, which is selected inside the tetrahedron such that its distance from each of the vertices is equal to unity, $OA = OT = OG = OC = 1$. It is well known that the center of gravity for a regular tetrahedron is three-quarter height below the highest point of the tetrahedron. This point is placed at the origin and the vertical height OA is taken to be one unit, as shown in Figure 2. Therefore the remaining vertical distance OP is one-third unit. From there, angle POC can be found by taking an inverse tangent that works out to be *70.5°*.

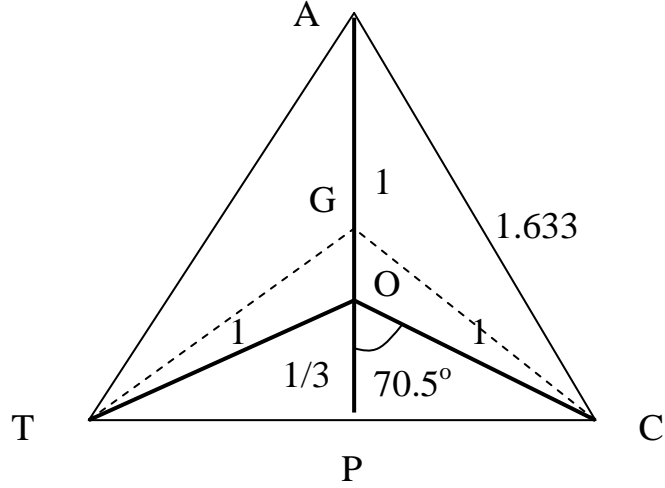

Figure 2: The tetrahedron

Using this information, the angle AOC is found to be *109.5°*. By some mathematical manipulation, the distance between any two nucleotide bases is the same and found to be *1.633*. This ensures an equal assignment of weightage for DNA four nucleotide bases. By this assignment, the distance between two points is range from *0* for a perfect match to *1.633* for a complete mismatch.

## C.    Vector-based representation of DNA motif

The conventional scalar-based scoring matrix [8, 9] can be converted to a vector-based scoring matrix as the following: For the given weightage of $p_b \in [0, 1]$ of the nucleotide base type $b \in \{A, T, G, C\}$, the vector-based scoring matrix is represented by a point at the coordinates $(x, y, z)$

$$x = \sqrt{(p_A A_x)^2 + (p_C C_x)^2 + (p_G G_x)^2 + (p_T T_x)^2} \qquad (1a)$$

$$y = \sqrt{\left(p_{\mathrm{A}}\mathrm{A}_y\right)^2 + \left(p_{\mathrm{C}}\mathrm{C}_y\right)^2 + \left(p_{\mathrm{G}}\mathrm{G}_y\right)^2 + \left(p_{\mathrm{T}}\mathrm{T}_y\right)^2} \qquad (1b)$$

$$z = \sqrt{\left(p_{\mathrm{A}}\mathrm{A}_z\right)^2 + \left(p_{\mathrm{C}}\mathrm{C}_z\right)^2 + \left(p_{\mathrm{G}}\mathrm{G}_z\right)^2 + \left(p_{\mathrm{T}}\mathrm{T}_z\right)^2} \, , \qquad (1c)$$

where $\vec{b} = \left(b_x,\, b_y,\, b_z\right)$ denotes the $x$, $y$ and $z$-coordinates of the nucleotide base type $b \in \{\mathrm{A,\,T,\,G,\,C}\}$ respectively.

## D.  Scalar-based scoring – Match alignment information

The distance between two points in the vector-based scoring matrix represents the degree of the similarity between them.  The nearer are two points, the better is the match.  This can be computed by using the dot product.  For the given coordinates $\left(x_n^s,\, y_n^s,\, z_n^s\right)$ of a DNA substring and $\left(x_n^m,\, y_n^m,\, z_n^m\right)$ of a motif at position $n$ of the same length $N$, the match alignment information is the value of the dot product of their coordinates as follows:

$$\text{Score of match alignment information} = \left| \sum_{n=1}^{N} \left[ \left(x_n^s,\, y_n^s,\, z_n^s\right) \bullet \left(x_n^m,\, y_n^m,\, z_n^m\right) \right] \right|. \qquad (2)$$

The resultant dot product gives a scalar.  The larger the number, the greater the similarity between the DNA substring and the motif.

## E.  Vector-based scoring – Mismatch alignment information

In this paper, a novel approach is proposed to analyze the types of mismatches based on mismatch alignment information. It can be used to decide whether one site has a greater chance of containing motifs over the others although they may have the same match alignment information. In normal double-stranded DNA, two-ring purine bases A and G are bigger in size as compared with one-ring pyrimidine bases T and C. One bigger two-ring nucleotide base pairs up with one smaller one-ring nucleotide base in a manner of A to T and G to C. DNA mismatches can generally be classified into three types: transitional, complementary and transversal mismatches.

(1)     Transition (transitional mismatch) occurs when one nucleotide base is replaced by another of the same size, *i.e.*, interchange between two-ring purine bases (A ↔ G) or between one-ring pyrimidine bases (T ↔ C). Because the replacement involves nucleotide bases of the same size, transitional mismatch is less likely to result in amino acid substitutions due to wobble, and is therefore more likely to persist as silent substitution in populations as single nucleotide polymorphisms. Hence, transitional mismatch is considered relatively acceptable.

(2)     Complement (complementary mismatch) occurs when one nucleotide base is replaced by its complementary nucleotide base of a different size, *i.e.*, interchange between complementary nucleotide bases (A ↔ T) or (G ↔ C).

(3)     Transversion (transversal mismatch) occurs when one nucleotide base is replaced by its non-complementary nucleotide base of a different size, *i.e.*, interchange between non-complementary nucleotide bases (A ↔ C) or (T ↔ G). Among three types of mismatches, transversal mismatch is considered the most significant. An alignment with a high number of transversal mismatches generally does not contain any motif.

For the given coordinates $\left(x_n^s, y_n^s, z_n^s\right)$ of a DNA substring and $\left(x_n^m, y_n^m, z_n^m\right)$ of a motif at position $n$ of the same length $N$, the mismatch alignment information is calculated by taking the cross product of their coordinates as follows:

$$\text{Score of mismatch alignment information} = \left\| \sum_{n=1}^{N} \left[\left(x_n^s, y_n^s, z_n^s\right) \times \left(x_n^m, y_n^m, z_n^m\right)\right] \right\|. \quad (3)$$

The resultant cross product gives a vector whose three components contain the following information:

Score of transitional mismatch alignment information

$$= \left| \sum_{n=1}^{N} \left[\left(x_n^s, y_n^s, z_n^s\right) \times \left(x_n^m, y_n^m, z_n^m\right)\right] \bullet \left(\vec{A} \times \vec{G} + \vec{C} \times \vec{T}\right) \right|; \quad (4a)$$

Score of complementary mismatch alignment information

$$= \left| \sum_{n=1}^{N} \left[\left(x_n^s, y_n^s, z_n^s\right) \times \left(x_n^m, y_n^m, z_n^m\right)\right] \bullet \left(\vec{A} \times \vec{T} + \vec{G} \times \vec{C}\right) \right|; \quad (4b)$$

Score of transversal mismatch alignment information

$$= \left| \sum_{n=1}^{N} \left[\left(x_n^s, y_n^s, z_n^s\right) \times \left(x_n^m, y_n^m, z_n^m\right)\right] \bullet \left(\vec{A} \times \vec{C} + \vec{T} \times \vec{G}\right) \right|. \quad (4c)$$

Here the respective mismatch vectors are shown in Table 1.

Table 1: Mismatch vectors

| Type | Pair | Coordinates | | |
|------|------|-------------|---|---|
|      |      | $x$ | $y$ | $z$ |
| **Transition** | $\vec{A} \times \vec{G}$ | $\dfrac{\sqrt{6}}{3}$ | $-\dfrac{\sqrt{2}}{3}$ | $0$ |
| | $\vec{C} \times \vec{T}$ | $-\dfrac{\sqrt{6}}{9}$ | $-\dfrac{\sqrt{2}}{3}$ | $-\dfrac{4\sqrt{3}}{9}$ |
| **Complement** | $\vec{A} \times \vec{T}$ | $0$ | $\dfrac{2\sqrt{2}}{3}$ | $0$ |
| | $\vec{G} \times \vec{C}$ | $\dfrac{2\sqrt{6}}{9}$ | $0$ | $-\dfrac{4\sqrt{3}}{9}$ |
| **Transversion** | $\vec{A} \times \vec{C}$ | $-\dfrac{\sqrt{6}}{3}$ | $-\dfrac{\sqrt{2}}{3}$ | $0$ |
| | $\vec{T} \times \vec{G}$ | $-\dfrac{\sqrt{6}}{9}$ | $\dfrac{\sqrt{2}}{3}$ | $-\dfrac{4\sqrt{3}}{9}$ |

## III.    RESULTS AND DISCUSSION

### 1.    Case study of TATA box in *Homo sapiens H4/g* gene

A TATA box is a segment of DNA sequence found in the promoter region of most genes in eukaryotes.  It is involved in the process of transcription by RNA polymerase, acting as the binding site of either transcription factor or histone [15]. Histone acts as a spool around which DNA winds and there are five major families, namely H1, H2A, H2B, H3 and H4.  H1 is known as the linker histone, while H2A, H2B, H3 and H4 are known as the core histones.  They are involved in the different stages of DNA packing.  In this case study, it is interesting to identify TATA box in *Homo sapiens H4/g* gene [16].  The *H4/g* gene is responsible for producing H4 histone.  By identifying the TATA box, it is possible to study the regulation of H4 histone production during DNA packing. As an example, a TATA box vector-based scoring matrix generated from RNA polymerase II promoter regions [17], as shown in the upper part of Table 2, is used to identify the possible sites containing the motif within *H4/g* gene. The corresponding weightage $p_b \in [0,1]$ of the nucleotide base type $b \in \{A, T, G, C\}$ and vector-based scoring matrix in terms of coordinates $(x, y, z)$ in equation (1) are shown in the middle and lower parts of Table 2, respectively. The vector-based scoring matrix is aligned with *H4/g* gene sequence base by base until the last nucleotide base. For each alignment, the match alignment information is calculated by using equation (2) and plotted in Figure 3.

Table 2: TATA box vector-based scoring matrix

|        | -2      | -1     | 0       | 1      | 2      | 3      | 4      | 5       |
|--------|---------|--------|---------|--------|--------|--------|--------|---------|
| **A**  | 16      | 352    | 3       | 354    | 268    | 360    | 222    | 155     |
| **T**  | 309     | 35     | 374     | 30     | 121    | 6      | 121    | 33      |
| **G**  | 18      | 2      | 2       | 5      | 0      | 20     | 44     | 157     |
| **C**  | 46      | 0      | 10      | 0      | 0      | 3      | 2      | 44      |
| $p_A$  | 0.0411  | 0.9049 | 0.0077  | 0.91   | 0.6889 | 0.9254 | 0.5707 | 0.3985  |
| $p_T$  | 0.7943  | 0.09   | 0.9614  | 0.0771 | 0.3111 | 0.0154 | 0.3111 | 0.0848  |
| $p_G$  | 0.0463  | 0.0051 | 0.0051  | 0.0129 | 0      | 0.0514 | 0.1131 | 0.4036  |
| $p_C$  | 0.1183  | 0      | 0.0257  | 0      | 0      | 0.0077 | 0.0051 | 0.1131  |
| $x$    | 0.9428  | 0      | 0.9428  | 0      | 0.4714 | 0      | 0.4714 | -0.2357 |
| $y$    | 0       | 0      | 0       | 0      | 0      | 0      | 0      | -0.4082 |
| $z$    | -0.3333 | 1      | -0.3333 | 1      | 0.3333 | 1      | 0.3333 | 0.3333  |

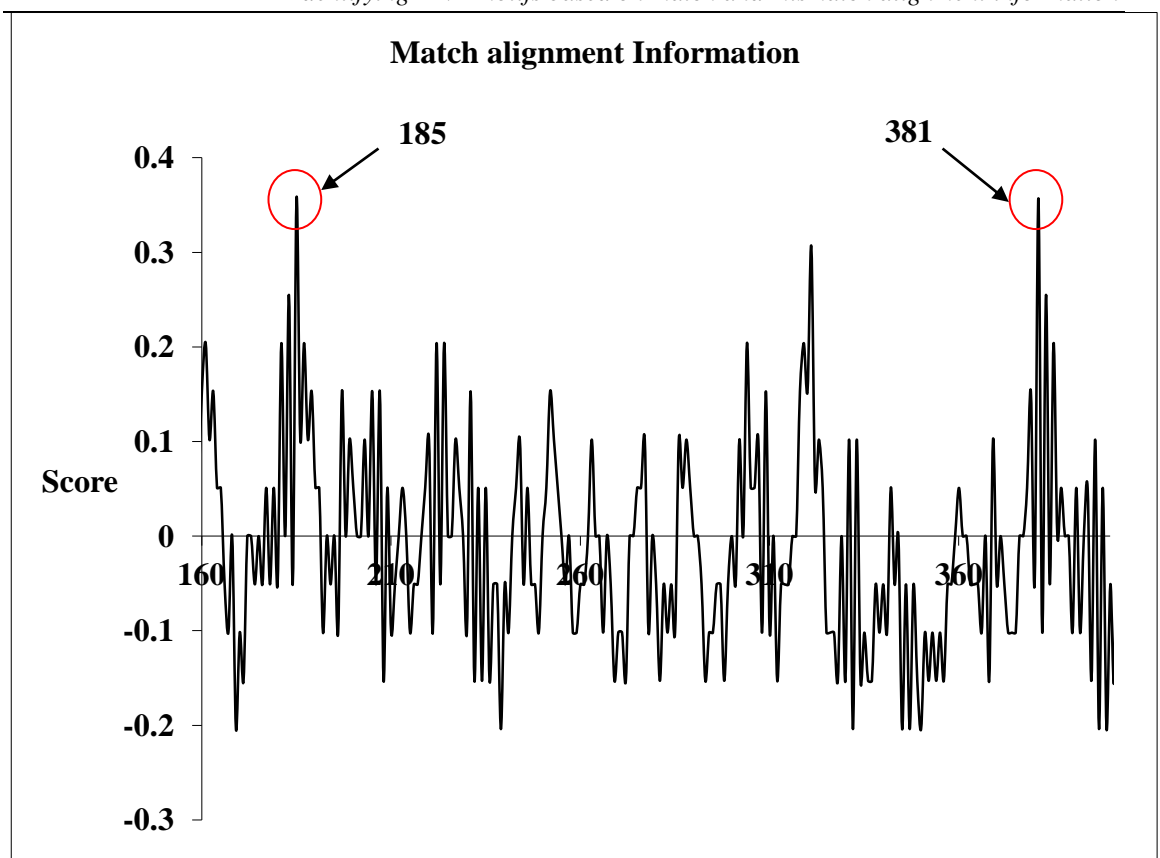

Figure 3: Match alignment information of *Homo sapiens H4/g* gene

## 2    Results and discussion

The plot composed of peaks and troughs.  Each peak represents a local high similarity between the TATA box vector-based scoring matrix and the regions within *H4/g* gene, and *vice versa* for troughs.  There are two highly possible sites (positions *185* and *381*) at which the TATA box could be situated.  However, it is not clear which is the real TATA box based on the match alignment information alone, since both have the same score.  In order to distinguish these two sites, the mismatch alignment information should be used.

The mismatch alignment information as discussed is classified into three categories: transition, complement and transversion.  Among these three types of mismatches, the transitional mismatch is considered the most acceptable. If the nucleotide base type A is replaced by G or *vice versa*, it is deemed to have less effect on the process of transcription [18]. This is because both nucleotide base types have the same size. However, if the different size nucleotide base type C replaces the nucleotide base type A in transversal mismatch, the degree of mismatch is greater than that of transitional one [18], because of having a significant effect on the transcription process. Standing in between is the complementary mismatch, although the nucleotide bases A and T may have a different size, they are complement as they bind to each other in a DNA chain. Given these three classes of mismatch alignment information, the most acceptable is the transitional mismatch. This is followed by the complementary mismatch and the least acceptable transversal mismatch.

By using equations (3) and (4), a comparison of the transitional mismatch alignment information shows that the scores are the same for both positions. It is still not possible to differentiate the dissimilarity between these two sites. Hence the complementary mismatch alignment information should be considered. Figure 4 shows that position *185* has the higher complementary mismatch alignment information as compared with position *381*. This implies that the nucleotide base A is replaced by the nucleotide base T, as well as G by C more often at position *185*. Hence the relatively-acceptable mismatch, the complementary mismatch, occurs at position *185*. Since the TATA box consists of a string of A and T bases, the complementary mismatch between A and T is unlikely to affect the function of the TATA box as a signal for transcription process. Therefore by using the complementary mismatch alignment information, position *185* is more likely to be the site for the TATA box in *H4/g* gene.

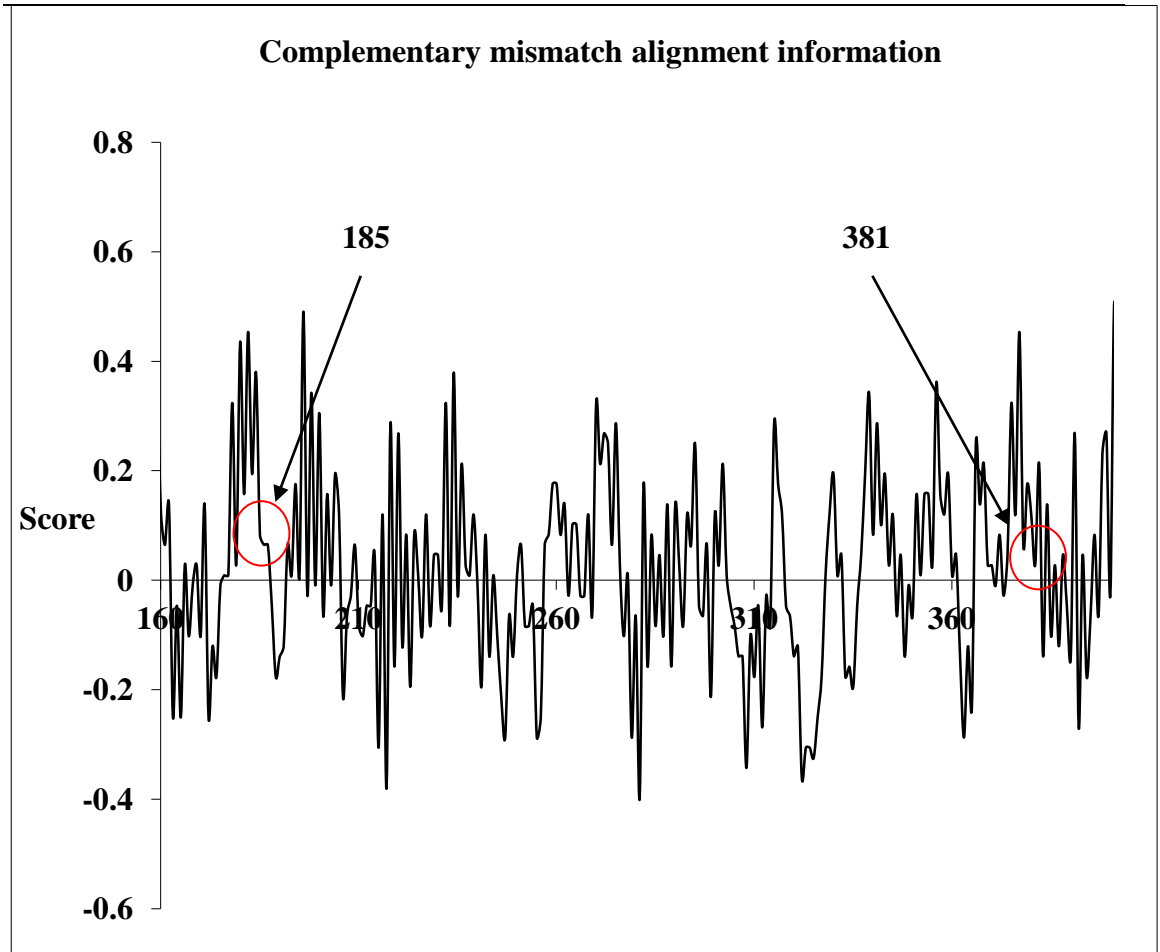

Figure 4: Complementary mismatch alignment information of *Homo sapiens H4/g*

gene

## IV.    CONCLUSION

A scalar-based scoring matrix is the useful way of representing an alignment and also used as a scoring means to predict motif sites.  However by using the match alignment information alone, there may be a high number of falsely predicted sites, especially for less conserved motifs.  In order to remove spurious sites, the mismatch alignment component should be considered.  The vector-based scoring matrix can be used to elaborate match and mismatch alignment information; match degree from the dot product and mismatch degree from the cross product. Using the mismatch alignment information, spurious sites can be removed and the efficiency in identifying real motifs can be improved [19].